\documentclass[12pt]{amsart}
\usepackage{amsfonts}

\def\cJ{{\mathcal J}}

\def\cF{{\mathcal F}}
\def\cA{{\mathcal A}}
\def\cM{{\mathcal M}}

\def\bR{{\mathbb R}}

\def\mbP{{\mathbb P}}
\def\mbF{{\mathbb F}}
\def\mbE{{\mathbb E}}

\newtheorem{theorem}{Theorem}
\theoremstyle{plain}

\newtheorem{definition}{Definition}

\newtheorem{lemma}[theorem]{Lemma}
\newtheorem{remark}[theorem]{Remark}

\numberwithin{equation}{section} \numberwithin{example}{section}
\numberwithin{definition}{section}

\renewcommand\endproof{\hfill $\Box$\vskip 0.15in}

\begin{document}
\title[Passive Scalar Equation]{Time Evolution of a Passive Scalar in a
Turbulent Incompressible Gaussian Velocity Field}
\author{S. V. Lototsky}
\curraddr[S. V. Lototsky]{Department of Mathematics, USC\\
Los Angeles, CA 90089}
\email[S. V. Lototsky]{lototsky@math.usc.edu}
\urladdr{http://math.usc.edu/$\sim$lototsky}
\thanks{The work was partially supported by the Sloan Research Fellowship and
by the ARO Grant DAAD19-02-1-0374.}
\author{B. L. Rozovskii}
\curraddr[B. L. Rozovskii]{Department of Mathematics, USC\\
Los Angeles, CA 90089} \email[B. L. Rozovskii]{rozovski@math.usc.edu}
\urladdr{http://www.usc.edu/dept/LAS/CAMS/usr/facmemb/boris/main.htm}
\thanks{The work was partially  supported by the ARO Grant DAAD19-02-1-0374.}
\today


\begin{abstract}
Passive scalar equation is considered in a turbulent homogeneous
incompressible Gaussian velocity field. The turbulent nature of the field
results in non-smooth coefficients in the equation. A strong, in the stochastic
sense, solution of the equation is constructed using the Wiener Chaos, and
the properties of the solution are studied. The results apply to both
viscous and conservative motions.
\end{abstract}

\maketitle

\section{Passive Scalar in a Gaussian Field}
\setcounter{theorem}{0}

\def\bv{{\mathbf v}}
We consider  the following transport equation  to
describe  the evolution  of a passive scalar $\theta$ in a random
velocity field $\bv$:
\begin{equation}
\label{eq:ps} \dot{\theta}(t,x)=0.5\nu\Delta \theta(t,x) - {\mathbf
v}(t,x) \cdot \nabla \theta(t,x)
 +f(t,x);\ x\in \bR^d,\ d>1.
\end{equation}
Our interest in this equation is motivated by the on-going   progress in the
study of the turbulent transport problem (E and Vanden Eijnden \cite{EVE},
 Gaw\c{e}dzki and Kupiainen \cite{GK}, Gaw\c{e}dzki and Vergasola \cite{GV},
  Kraichnan \cite{Krch}, etc.)

We assume in (\ref{eq:ps})
 that $\bv=\bv(t,x)\in \bR^d$, $d\geq 2$,  is an isotropic Gaussian
vector field with zero mean and covariance
$$
E(v^i(t,x)v^j(s,y))=\delta(t-s)C^{ij}(x-y)
$$
with some matrix-valued function $C=(C^{ij}(x), i,j=1, \ldots,
d)$.
 It is well-known (see, for example,  LeJan \cite{LeJan}) that in
the physically interesting models the matrix-valued function
$C=C(x)$ has the Fourier transform $\hat{C}=\hat{C}(z)$ given by
$$
\hat{C}(z)=\frac{A_0}{(1+|z|^2)^{(d+\alpha)/2}}\left(a\frac{zz^*}{|z|^2}+\frac{b}{d-1}
\left(I-\frac{zz^T}{|z|^2}\right) \right),
$$
where $z^*$ is the row vector $(z_1, \ldots, z_d)$, $z$ is the
corresponding column  vector, $|z|^2=z^*z$, $I$ is the identity
matrix; $\alpha>0,a\geq 0,b\geq 0,A_0>0$ are real numbers. Similar
to \cite{LeJan},
 we assume that $0<\alpha<2$.

By direct computation (cf. \cite{BH}),
 the vector field  $\bv=(v^1, \ldots, v^d)$ can be written as
\begin{equation}
\label{eq:v} v^i(t,x)=\sum_{k\geq 0} \sigma^i_k(x)\dot{w}_k(t),
\end{equation}
where $\dot{w}_k(t)$, $k\geq 1$, are independent standard
Gaussian white noises and
$\{\sigma_k, \ k\geq 1\}$ is a CONS in the space $H_C$, the
reproducing kernel Hilbert space corresponding to the kernel
function $C$. The space $H_C$ is all or part of the Sobolev space
$H^{(d+\alpha)/2}(\bR^d; \bR^d)$. It follows from (\ref{eq:v}) that
 $\sum_k \sigma_{k}^i(x)\sigma_{k}^j(y)=C^{ij}(x-y)$ for all $x,y$;
 in particular, $\sigma_k^i(x)\sigma_k^j(x)=C^{ij}(0)$ for all $x$.

If $a>0$ and $ b>0$, then the matrix $\hat{C}$ is invertible and
$$
H_C=\{ f\in \bR^d: \int_{\bR^d}
\hat{f}^*(z)\hat{C}^{-1}(z)\hat{f}(z)dz < \infty \} =
H^{(d+\alpha)/2}(\bR^d; \bR^d),
$$
because $\|\hat{C}(z)\| \sim (1+|z|^2)^{-(d+\alpha)/2}$.

If $a>0$ and $b=0$, then
$$
H_C=\left\{ f\in \bR^d: \int_{\bR^d}
|\hat{f}(z)|^2(1+|z|^2)^{(d+\alpha)/2}dz < \infty; \
zz^*\hat{f}(z)=|z|^2\hat{f}(z) \right\},
$$
the subset of gradient fields in $H^{(d+\alpha)/2}(\bR^d; \bR^d)$
(those are vector fields $f$ for which $\hat{f}(z)=z\hat{F}(z)$
for some scalar $F \in H^{(d+\alpha+1)/2}$).

If $a=0$ and $b>0$, then
$$
H_C=\left\{ f\in \bR^d: \int_{\bR^d}
|\hat{f}(z)|^2(1+|z|^2)^{(d+\alpha)/2}dz < \infty; \
z^*\hat{f}(z)=0 \right\},
$$
the subset of divergence free fields in $H^{(d+\alpha)/2}(\bR^d;
\bR^d)$.

By the embedding theorems, each  $\sigma_k^i$ is a bounded
continuous function on $\bR^d$; in fact, every  $\sigma_k^i$ is
H\"{o}lder continuous of order $\alpha/2$. In addition, being an
element of the corresponding space $H_C$,
 each $\sigma_k$ is a gradient field if $b=0$ and is divergence free if
 $a=0$.

 To simplify the further presentation and to make the model (\ref{eq:ps})
  more physically relevant,  we consider the divergence-free  velocity field
  and assume that the stochastic integration
  is in the sense of Stratonovich. Under these assumptions,
  equation (\ref{eq:ps}) becomes
\begin{equation}
\label{eq:ps2s}
d{\theta}(t,x)=0.5\nu\Delta \theta(t,x)dt-
\sum_{k}\sigma_{k}(x)\cdot \nabla\theta(t,x) \circ dw_k(t).
\end{equation}
With  divergence-free functions $\sigma_k$, the equivalent Ito formulation
is
\begin{equation}
\label{eq:ps2i}
d{\theta}(t,x)=0.5(\nu\Delta \theta(t,x)+C^{ij}(0)D_iD_j\theta(t,x))dt-
\sigma_{k}^i(x)D_i\theta(t,x) dw_k(t).
\end{equation}
In what follows, we  construct a solution of (\ref{eq:ps2i}) using Wiener Chaos.

\section{A Review of the Wiener Chaos}

Let $\mbF=(\Omega, \cF, \{\cF_t\}_{t\geq 0}, \mbP)$ be a stochastic
basis with the usual
assumptions.
  On $\mbF$ consider a collection $(w_k(t),k\geq 1,t\geq 0)$ of
 independent standard Wiener processes. For a fixed $0<T<\infty$,
let  $\cF^W_T$  be the sigma-algebra generated by $w_k(t),\ k\geq 1, \ 0<t<T$,
 and  $L_2(\cF^W_T)$
the collection of $\cF^W_T$-measurable square integrable random variables.

For   the Fourier cosine  basis $\{m_k,\ k\geq 1,\}$ in $L_2((0,T))$ with
\begin{equation}
\label{eq:cosbas}
m_1(t)=\frac{1}{\sqrt{T}}, \
m_k(t)=\sqrt{\frac{2}{T}}\cos\left(\frac{\pi(k-1)t}{T}\right),\  k\geq 2,
\end{equation}
define the independent
standard Gaussian random variables
$$
\xi_{ik}=\int_0^T m_i(s)dw_k(s).
$$

Consider the collection of multi-indices
 $$
 \cJ=\Big\{ \alpha =(\alpha_i^k,\ i,k\geq 1),\ \alpha_{i}^k\in
\{0,1,2,\ldots\},\ \sum_{i,k} \alpha_i^k<\infty \Big\}.
$$
The set $\cJ$ is countable, and, for every $\alpha \in \cJ$, only
finitely many of $\alpha_i^k$ are not equal to zero.
For $\alpha \in {\mathcal J}$, define
$$
|\alpha|=\sum_{i,k} \alpha_i^k, \ \alpha!=\prod_{i,k}\alpha_i^k!,
$$
and
$$
\xi_{\alpha}=\frac{1}{\sqrt{\alpha!}}\prod_{i,k}H_{\alpha_{i}^{k}}(\xi_{ik}),
$$
where
$$
 H_{n}(t)=e^{t^{2}/2}
\frac{d^{n}}{dt^{n}}e^{-t^{2}/2}
$$
is $n$-th Hermite polynomial. In particular, if $\alpha\in\cJ$ is such that
$\alpha_i^k=1$ if  $i=j$ and $k=l$, and $\alpha^k_i=0$ otherwise,
 then  $\xi_{\alpha}=\xi_{jl}$.

\begin{definition} The  space $L_2(\cF^W_T)$ is called the Wiener Chaos space.
The $N$-th Wiener Chaos is the linear subspace of $L_2(\cF^W_T)$, generated by
$\xi_{\alpha},\ |\alpha|=N$.
\end{definition}

The following is  a classical results of Cameron and Martin \cite{CM}.
\begin{theorem}
\label{th:CM}
The collection $\{ \xi_{\alpha},\ \alpha \in \cJ\}$ is  an orthonormal basis
in the space $L_2(\cF^W_T)$.
\end{theorem}
In addition to the original source \cite{CM}, the proof of this theorem can be
 found in many other places, for example, in  \cite{HKPS}.
By Theorem \ref{th:CM} every element $v$ of $L_2(\cF^W_T)$ can be written as
$$
v=\sum_{\alpha \in \cJ} v_{\alpha}\xi_{\alpha},
$$
where
$v_{\alpha}=\mbE (v\xi_{\alpha})$.

 \section{The Wiener Chaos Solution of the Passive Scalar Equation}

 Using the summation convention,
 define the operators  $\cA=0.5(\nu\Delta+C^{ij}(0)D_iD_j)$ and $\cM_k=\sigma_k^iD_i$.
 Assume that $\theta_0 \in L_2(\bR^d)$ and
 define $\theta_{\alpha}(t,x)$ by
\begin{equation}
\label{eq:def2}
\theta_{\alpha}(t)=\theta_{0}I(|\alpha|=0) + \int_0^t \cA \theta_{\alpha}(s,x)ds+
\int_0^t\sum_{i,k}\sqrt{\alpha_i^k}\cM_k \theta_{\alpha^-(i,k)}(s)m_i(s)ds.
\end{equation}
Notice that for every function $f \in H^1_2(\bR^d)$,
$$
\sum_{k\geq 1} \|\cM_k f\|^2_{L_2(\bR^d)}=(\sigma_k^j\sigma_k^iD_if, D_jf)=
(C^{ij}(0)D_if, D_jf),
$$
where $(\cdot, \cdot)$ is the inner product in $L_2(\bR^d)$.
Since the matrix $(C^{ij}(0), i,j=1, \ldots, d)$ is positive definite,
we conclude that there exist positive numbers $c_1, c_2$ so that, for
every function $f \in H^1_2(\bR^d)$,
\begin{equation}
\label{eq:PSnorm}
c_1\|\nabla f\|_{L_2(\bR^d)}^2 \leq\sum_{k\geq 1} \|\cM_k f\|^2_{L_2(\bR^d)}
\leq c_2 \|\nabla f\|_{L_2(\bR^d)}^2.
\end{equation}

\begin{theorem}
\label{th:main}
\begin{enumerate}
\item For every $\nu\geq 0$ and every $t\in [0,T]$, the series
\begin{equation}
\label{eq:PSsum}
\sum_{\alpha \in \cJ} \theta_{\alpha}(t,x)\xi_{\alpha}
\end{equation}
converges in $L_2(\Omega; L_2(\bR^d))$ to a process
$\theta=\theta(t,x)$.
\item If $\nu>0$, then, for every $\varphi\in C^{\infty}_0(\bR^d)$,
  the process $\theta(t,x)$ satisfies
\begin{equation}
\label{eq:PSsol1}
\begin{split}
(\theta,\varphi)(t)&=(\theta_0, \varphi)-0.5\nu\int_0^t
(\nabla \theta,\nabla \varphi)(s)ds-
0.5\int_0^tC^{ij}(0)(D_i\theta, D_j\varphi)(s)ds\\
&-\int_0^t(\sigma^i_kD_i\theta,\varphi)dw_k(s)
\end{split}
\end{equation}
with probability one for all $t \in [0,T]$ at once, where
$(\cdot, \cdot)$ is the inner product in $L_2(\bR^d)$.
Also,
\begin{equation}
\label{eq:PSnorm1}
\mbE \|\theta\|_{L_2(\bR^d)}^2(t) +
\nu \int_0^t\mbE\|\nabla\theta\|_{L_2(\bR^d)}^2(s)ds
= \|\theta_0\|_{L_2(\bR^d)}^2.
\end{equation}
\item If $\nu=0$, then, for every $\varphi\in C^{\infty}_0(\bR^d)$,
  the process $\theta(t,x)$ satisfies
\begin{equation}
\label{eq:PSsol0}
(\theta,\varphi)(t)=(\theta_0, \varphi)+0.5\int_0^t
C^{ij}(0)(\theta, D_iD_j\varphi)(s)ds
+\int_0^t(\theta,\sigma^i_kD_i\varphi)dw_k(s)
\end{equation}
with probability one for all $t \in [0,T]$ at once.
Also,
\begin{equation}
\label{eq:PSnorm0}
\mbE \|\theta\|_{L_2(\bR^d)}^2(t)
\leq  \|\theta_0\|_{L_2(\bR^d)}^2.
\end{equation}
\end{enumerate}
\end{theorem}

\begin{remark} Equalities (\ref{eq:PSsol1}) and (\ref{eq:PSsol0}) mean that
  $\theta=\theta(t,x)$ is the solution of the transport equation in the traditional
  sense of theory of stochastic partial differential equations, that is, it is
  a strong solution in the stochastic sense, satisfying the corresponding equation
  in the generalized function sense. The solution is also unique in the
  class of $L_2((0,T)\times \Omega; L_2(\bR^d)) $ random functions, because any
  other solution will automatically have the same Wiener Chaos expansion.
  The uniqueness can, in fact, be established in a much wider class of generalized
  random functions.
  \end{remark}

The proof of Theorem \ref{th:main} is based on the following lemmas.

\begin{lemma}
\label{lm:LMR}
The system of equations (\ref{eq:def2}) has a unique solution so that
every $\theta_{\alpha}$ is a smooth bounded function of $x$ for $t>0$ and,
 if  $T_t, \ t\geq 0,$ is the heat semigroup generated by the operator
 $0.5(\nu\Delta+C^{ij}(0)D_iD_j)$,
then, for every $N\geq 0$,
\begin{equation}
\label{eq:L2norm}
\begin{split}
&\sum_{|\alpha|=N} |\theta_\alpha(t,x)|^2
\\
 &=\sum_{k_1, \ldots, k_N=1}^{\infty} \int_0^t \int_0^{s_N}\ldots\int_0^{s_2}
 |T_{t-s_N}\cM_{k_N} \ldots T_{s_2-s_1}\cM_{k_1}T_{s_1}\theta_0(x)|^2ds_1\ldots ds_N.
 \end{split}
 \end{equation}
 and
\begin{equation}
\label{eq:L2gnorm}
\begin{split}
&\sum_{|\alpha|=N} |\nabla\theta_\alpha(t,x)|^2
\\
 &=\sum_{k_1, \ldots, k_N=1}^{\infty} \int_0^t \int_0^{s_N}\ldots\int_0^{s_2}
 |\nabla T_{t-s_N}\cM_{k_N}
  \ldots T_{s_2-s_1}\cM_{k_1}T_{s_1}\theta_0(x)|^2ds_1\ldots ds_N.
 \end{split}
 \end{equation}
 \end{lemma}
  {\bf Proof.} See Proposition A.1 in \cite{LMR}.
  \endproof

\begin{lemma}
\label{lm:tail1}
Assume that $\nu\geq 0$.
Define $\theta_N(t,x)=\sum_{n=0}^N \sum_{|\alpha|=n}\theta_{\alpha}(t,x)\xi_{\alpha}$.
Then, for all $t \in [0,T]$,
\begin{equation}
\label{eq:PStr}
\begin{split}
&\mbE \|\theta_N\|^2_{L_2(\bR^d)}(t)=\|\theta_0\|_{L_2(\bR^d)}^2-
\nu\sum_{n=0}^N \sum_{|\alpha|=n}
\int_0^t\|\nabla \theta_{\alpha}\|_{L_2(\bR^d)}^2(s)ds
\\
&-\sum_{k_1, \ldots, k_{N+1}} \int_0^t\ldots\int_0^{s_2}
 \|\cM_{k_{N+1}}T_{s-s_N}\cM_{k_N}
 \ldots T_{s_2-s_1}\cM_{k_1}T_{s_1}\theta_0\|_{L_2(\bR^d)}^2ds^Nds.
\end{split}
\end{equation}
\end{lemma}
{\rm {\bf Proof.} By Lemma \ref{lm:LMR}, after
integration with respect to $x$,
\begin{equation}
\label{eq:PSL2norm}
\begin{split}
&\sum_{|\alpha|=N} \|\theta_\alpha\|_{L_2(\bR^d)}^2(t)
\\
 &=\sum_{k_1, \ldots, k_{N}=1}^{\infty} \int_0^t \int_0^{s_N}\ldots\int_0^{s_2}
 \|T_{t-s_N}\cM_{k_N} \ldots
 T_{s_2-s_1}\cM_{k_1} T_{s_1}\theta_0\|_{L_2(\bR^d)}^2ds_1\ldots ds_N.
 \end{split}
 \end{equation}
  If $F_N(t)=\sum_{|\alpha|=N} \|\theta_\alpha(t)\|_{L_2(\bR)}^2$, then
\begin{equation}
\begin{split}
&\frac{d}{dt}F_N(t)
\\
&=\sum_{k_1, \ldots, k_{N}}
\int_0^t \int_0^{s_{N-1}}\ldots\int_0^{s_2}
 \|\cM_{k_N} T_{t-s_{N-1}}\cM_{k_{N-1}} \ldots
 T_{s_2-s_1}\cM_{k_1} T_{s_1}u_0\|_{L_2(\bR)}^2ds^{N-1}
 \\
 &+2\sum_{k_1, \ldots, k_{N}}
 \int_0^t\ldots\int_0^{s_2}
 \left(\cA T_{t-s_N}\cM \ldots
  T_{s_1}u_0, T_{t-s_N}\cM \ldots
 T_{s_2-s_1}\cM_{k_N} T_{s_1}u_0\right) ds^N.
 \end{split}
 \end{equation}
 It remains to notice that, for every smooth function $f=f(x)$,
 $$
 2(\cA f,f)=-\nu\|\nabla f\|_{L_2(\bR)}^2-
 \sum_{k\geq1}\|\cM_k f\|_{L_2(\bR)}^2.
$$
 Equality (\ref{eq:PStr}) now follows.
 \endproof
Notice that (\ref{eq:PStr}) implies both the $L_2(\Omega;L_2(\bR^d))$
convergence of the
series
$$
\sum_{\alpha}\theta_{\alpha}(t,x)\xi_{\alpha}
$$
 for every $t\in [0,T]$ and
inequality (\ref{eq:PSnorm0}).

 \begin{lemma}
 \label{lm:tail2}
If $\nu>0$, then, for every $t\in [0,T]$,
\begin{equation}
\label{eq:PStail}
\lim_{N\to \infty}
\sum_{k_1, \ldots, k_{N+1}} \int_0^t\ldots\int_0^{s_2}
 \|\cM_{k_{N+1}}T_{s-s_N}\cM_{k_N}
 \ldots T_{s_2-s_1}\cM_{k_1}T_{s_1}\theta_0\|_{L_2(\bR^d)}^2 ds^Nds=0.
 \end{equation}
 \end{lemma}
 {\bf Proof.}
Define
\begin{equation}
\label{eq:PStail1}
\begin{split}
&F_N(t)
\\
&= \sum_{k_1, \ldots, k_{N+1}}
\int_0^t\ldots\int_0^{s_2}
 \|\cM_{k_{N+1}}T_{s-s_N}\cM_{k_N}
 \ldots T_{s_2-s_1}\cM_{k_1}T_{s_1}\theta_0\|_{L_2(\bR^d)}^2 ds^Nds.
 \end{split}
 \end{equation}
 By (\ref{eq:PSnorm}) and Lemma \ref{lm:LMR},
 $F_N(t)\leq c_2 \sum_{|\alpha|=N} \int_0^t
 \|\nabla \theta_{\alpha}\|_{L_2(\bR^d)}^2(s)ds$. Lemma \ref{lm:tail1} then implies that
 the series $\sum_{N\geq 0} F_N(t)$ converges for all $t \in [0,T]$. Therefore,
 $\lim_{N\to \infty} F_N(t)=0$ and the statement of the lemma follows.

 \endproof

 Since (\ref{eq:PStr}) and (\ref{eq:PStail}) imply (\ref{eq:PSnorm1}),
 to complete the proof of the theorem  it remains to establish (\ref{eq:PSsol1})
 and (\ref{eq:PSsol0}).  The necessary arguments are similar to  the proof
 of Theorem 3.5 in \cite{MR.gs}

}

\providecommand{\bysame}{\leavevmode\hbox to3em{\hrulefill}\thinspace}
\providecommand{\MR}{\relax\ifhmode\unskip\space\fi MR }
\providecommand{\MRhref}[2]{%
  \href{http://www.ams.org/mathscinet-getitem?mr=#1}{#2}
}
\providecommand{\href}[2]{#2}

\end{document}